\documentclass{aa}

\usepackage{txfonts}

\usepackage{graphicx}
\usepackage[colorlinks]{hyperref}

\hypersetup{
    linkcolor=blue,
    citecolor=blue,
    filecolor=magenta,      
    urlcolor=blue
}

\def\say#1{``#1''}

\newcommand{\Msun}{\mbox{$\mathrm{M}_{\odot}$}}

\newcommand{\Teff}{\mbox{$T_{\mathrm{eff}}$}}

\begin{document}

\title{Lithium evolution in the Galactic thin disc from Main-Sequence and early Red-Giant-Branch stars}
\titlerunning{Lithium evolution in the Galactic thin disc from MS and early RGB stars}

\subtitle{}

   \author{C. T. Nguyen
        \inst{1,2}
        \and
        G. Cescutti   
        \inst{1,2}
        \and
        F. Matteucci
        \inst{1,2,3}
        \and
        F. Rizzuti
        \inst{4,2,5}
        \and
        A. Mucciarelli
        \inst{6,7}
        \and
        D. Romano
        \inst{6}
        \and
        L. Magrini
        \inst{8}
        \and
        A. J. Korn         
        \inst{9}
        \and
        A. Bressan
        \inst{10}
        \and
        L. Girardi
        \inst{11}
          }

        \institute{
        University of Trieste, Piazzale Europa, 1, Trieste, Italy,\\
        \email{chi.nguyen@inaf.it}
        \and
        INAF Osservatorio Astronomico di Trieste, Via Giambattista Tiepolo, 11, Trieste, Italy
        \and
        Institute for Fundamental Physics of the Universe, via Beirut, 2, 34151 Trieste, Italy
        \and
        Heidelberger Institut für Theoretische Studien, Schloss-Wolfsbrunnenweg 35, D-69118 Heidelberg, Germany
        \and
        INFN, Sezione di Trieste, via Valerio 2, I-34134 Trieste, Italy
        \and
        INAF, Osservatorio di Astrofisica e Scienza dello Spazio, Via Gobetti 93/3, 40129 Bologna, Italy
        \and
        Dipartimento di Fisica e Astronomia, Università degli Studi di Bologna, Via Gobetti 93/2, 40129 Bologna, Italy
        \and
        INAF, Osservatorio Astrofisico di Arcetri, Largo E. Fermi 5, 50125 Firenze, Italy
        \and
        Division of Astronomy and Space Physics, Department of Physics and Astronomy, Uppsala University, Box 516, SE-75120 Uppsala, Sweden
        \and
        SISSA, Via Bonomea 265, I-34136 Trieste, Italy
        \and
        INAF - Osservatorio Astronomico di Padova, Vicolo dell'Osservatorio 5, Padova, Italy
      }

\authorrunning{Nguyen et al.}

   \date{}

\abstract {The role of novae as producers of galactic lithium-7 (Li) has been suggested since the 1970s, and it has been reconsidered recently with the detection of beryllium-7 in their outbursts. 
At the same time, stellar models are moving forward to comprehend the discrepancy between the primordial Li abundance predicted by the standard Big Bang Nucleosynthesis theory and the measured value of old dwarf stars. 
In this work, we follow the evolution of Li in the galactic thin disc starting from a primordial value of A(Li)=2.69 dex and applying Li depletion corrections of the stellar model with overshoot to our chemical evolution models. We use the upper envelope of the observational data to constrain the models. In addition to the dwarf main sequence (MS) stars, our analysis includes, for the first time, the early red-giant-branch (RGB) stars. Besides the renowned Spite plateau of the MS stars at low metallicities, we also confirm the existence of a second A(Li) plateau of the early RGB stars, which can be explained by our model with the corrections from stellar models. Our best-fit model is obtained with an effective averaged Li yield $^{Li}Y_\mathrm{Nova}=2.34\times 10^{-5}\Msun$ during the whole lifetime of a nova. This reinforces the possibility that novae are the main galactic Li source, together with the stellar models' ability to comprehend the \say{cosmological Li problem} in this context.}

   
   \maketitle
   
\section{Introduction}

Tracing galactic lithium-7 (Li) provides abundant information on the evolution of the Milky Way \citep[for example,][and references therein]{1995A&A...303..460M,2001ApJ...559..909T,2012A&A...542A..67P}. Most of the models included novae as primary Li factory \citep[][]{1991A&A...248...62D,1999A&A...352..117R}, but for a long time there were no observational evidences supporting Li production in nova outbursts. 

Since the discovery of \citet{2015Natur.518..381T} where Be was detected in the classical Nova Delphini 2013, classical nova explosions were confirmed to be the main source of the galactic Li production among other possible sources such as red giant or asymptotic giant branch (AGB) stars. 
Briefly after that, \citet{2015ApJ...808L..14I} reported the detection of the Li \,\texttt{I} at $\lambda=\,$6708~$\AA$ line in the spectra of Nova Centauri 2013 (V1369Cen). 
This was further supported by the detection of the isotope Be in many other novae \citep[][]{2016MNRAS.463L.117M,2016ApJ...818..191T,2018MNRAS.478.1601I,2020MNRAS.492.4975M,2021ApJ...916...44A,2022MNRAS.509.3258M,2023MNRAS.518.2614M}. 
Currently, a possible detection of a 478 keV emission line emitted in the Be decay into Li has been reported by \citet{2025A&A...698A.291I} for V1369Cen. 
This is due to the short-lived lifetime of Be ($\sim 53$ days) and its decay into Li. 
Indeed, the amount of ejected Li per nova event, $\approx 1-10\times 10^{-9}\Msun$, was determined in these cases. 
This strongly indicates that the main source of Li production in the Milky Way is from nova systems. 

Over the last decade, observational surveys have provided a huge amount of data on Li abundances, from galactic discs to the bulge, from star clusters to halo field stars, from the main-sequence (MS) to the advanced red-giant and asymptotic-giant branch evolutions. For example, the Gaia-ESO survey \citep[][]{2018A&A...610A..38F,2020A&A...640L...1R,2021A&A...651A..84M, 2021A&A...653A..72R}, the Galactic Archaeology with HERMES survey \citep[GALAH,][]{2015MNRAS.449.2604D,2024MNRAS.528.5394W}, the Large Sky Area Multi-Object Fiber Spectroscopic Telescope survey \citep[LAMOST,][]{2019ApJS..245...33G,2024ApJS..271...58D}, and the upcoming 4MOST MIlky way Disc And BuLgE High-Resolution survey \citep[4MIDABLE-HR,][]{2019Msngr.175...35B}. 
These surveys are a great source of observational data for chemical evolution models to study precisely the properties of galactic Li evolution. 

In particular, various models recently claim the importance of novae for the enrichment of Li in the solar neighbourhood, for example, \citet{2019MNRAS.482.4372C}, \citet{2021A&A...653A..72R}, \citet{2024ApJ...971....4G}, and \citet{2024A&A...691A.142B}.
These works agree that novae are the main source of the galactic Li production. However, the precise amount of Li yield per nova event differs between different works. In particular, \citet{2019MNRAS.482.4372C} claimed an amount of $1.8\times 10^{-9}\Msun$ to be the best constrained value (assuming about $10^4$ explosion events occur during the whole lifetime of novae), while \citet{2024A&A...691A.142B} claimed a $\sim$3 times larger value, $5\times 10^{-9}\Msun$ with their one-zone model (assuming an average mass of the ejecta of $5\times 10^{-5}\Msun$ in a novae lifetime). Furthermore, the dependence of nova Li yield on other physical parameters such as metallicity, ejecta mass, and delay time gives rise to the complexity of nova treatment in galactic chemical evolution model \citep[see][and references therein for more details]{2022ApJ...933L..30K, 2024A&A...689A.222K}.

On the other hand, observations of metal-poor dwarf-MS stars ($\mathrm{[Fe/H]}\leq -1$) indicate an A(Li) plateau commonly known as the Spite plateau \citep[$\mathrm{A(Li)}\approx 2.2$ dex;][]{1982A&A...115..357S,1982Natur.297..483S,2002A&A...395..515B,2005A&A...442..961C,2007ApJ...671..402K}. This measured value is $\sim 3$ times smaller than the predicted primordial value which is obtained from the Big Bang Nucleosynthesis studies \citep[$\mathrm{A(Li)}\approx 2.7$ dex;][]{2014JCAP...10..050C,2016RvMP...88a5004C,2019JETP..128..707S,2020ApJ...901..127I,2021MNRAS.502.2474P,2024APh...16202995S} assuming the cosmological parameters determined by Planck \citep[][]{2014A&A...571A..16P}. 
The origin of this discrepancy is still debated, and is usually known as the \say{cosmological Li problem}. For example, \citet{2012A&A...539A..70E}, \citet{2015MNRAS.452.3256F} and \citet{2025A&A...696A.136N} proposed different mechanisms involving Li depletion during the PMS evolution, while \citet{2005ApJ...619..538R} moderated atomic diffusion with turbulent mixing during the MS. Recently, \citet{2025A&A...696A.136N} presented a grid of stellar models in which the envelope overshoot efficiency parameter was modified during the PMS evolution of the stars to better understand this issue. Their models were calibrated with the well-studied globular cluster NGC 6397. Taking advantage of the provided stellar grid, we explored the amount of stellar Li depletion relative to the primordial value ($\mathrm{A(Li)}=2.69$ dex), and coupled it with the predictions of our chemical evolution code. For the purpose of this work, we adopt the simplified treatment of nova Li yield from \citet{2019MNRAS.482.4372C} and introduce a correction term accounting for stellar Li depletion due to the evolution of stars.

Moreover, \citet{2022A&A...661A.153M} recently discovered a thin plateau among the metal-poor red-giant-branch (RGB) stars. Their sample contains 58 early RGB stars with metallicity ranges $-7\leq\mathrm{[Fe/H]}\leq -1.3$, and high-resolution spectra were used for the derivation of A(Li). The authors discovered a plateau for these early RGB metal-poor stars, with the mean value $\mathrm{A(Li)}=1.09\pm 0.08$ dex, that exhibits a trend similar to that of MS stars \citep[or the so-called Spite plateau;][]{1982A&A...115..357S,1982Natur.297..483S}. 
Hence, in this work, we also study the behaviour of A(Li) at this early RGB evolution to investigate this discovered plateau and the role of nova production in Li enrichment at this evolutionary stage.  
For this purpose, we selected two subsamples of dwarf-MS and early-RGB stars from the recent catalogues of GALAH DR4 \citep[][]{2024MNRAS.528.5394W} and Gaia-ESO \citep[][]{2021A&A...651A..84M,2021A&A...653A..72R}, complemented with the sample of F- and G-type dwarf-MS stars from \citet{2018A&A...615A.151B}. 

The paper is organised into six sections. In Sect.~\ref{sample_selection}, we describe the methodology to select our data samples from the big surveys GALAH and Gaia-ESO. Section.~\ref{stellar_model} shows the predictions from the stellar models regarding the depletion of $\mathrm{A(Li)}$ relative to the primordial value. In Sect.~\ref{cem}, we describe the chemical evolution model that is used for this work. We present the obtained results in Sect.~\ref{results} and draw our conclusions in Sect.~\ref{conclusion}.

\section{Sample selection}\label{sample_selection}

\begin{figure}[t]
    \centering
    \includegraphics[width=\columnwidth]{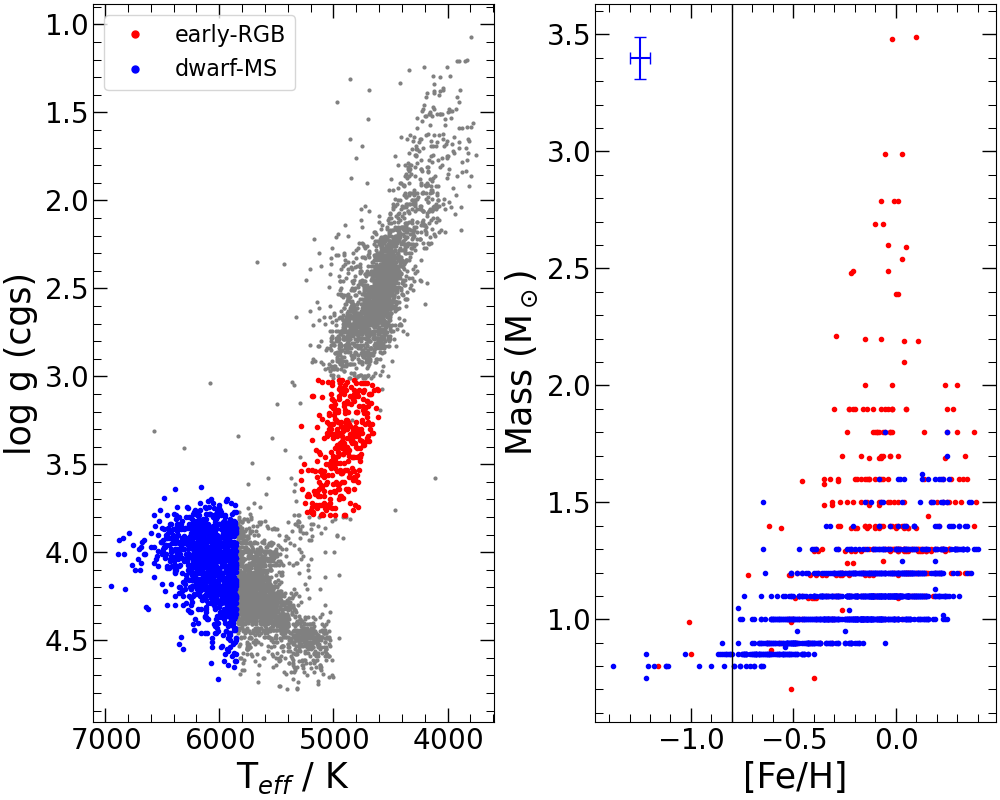}
    \caption{Left panel: the sample of 7\,347 field stars from \citet{2021A&A...651A..84M}, with our selected subsamples highlighted in red and blue. Right panel: the mass distribution with metallicity ([Fe/H]) of stars in our selected samples. The blue error bar indicates the mean uncertainty value of our selected dwarf-MS sample. The black vertical line marks $\mathrm{[Fe/H]}=-0.8$ dex.}
    \label{Kiel_GaiaESO}
\end{figure}

The catalogue of \citet{2021A&A...651A..84M} presents high-quality data from the Gaia-ESO iDR6 survey. It contains 7\,347 field stars with metallicity in the range $-1\leq \mathrm{[Fe/H]}\leq 0.5$ dex. The Li abundances are derived with the one-dimensional (1D) local thermodynamic equilibrium (LTE) assumption \citep[][]{2022A&A...668A..49F}. We selected two subsamples of dwarf-MS stars ($\log g> 3.6$ and $T_\mathrm{eff}\geq 5850$ K) and early RGB stars ($3.1\leq\log g\leq 3.8$ and $4600\leq T_\mathrm{eff}/\mathrm{K}\leq 5300$) for our analysis in this work. 
The choice of $\log g$ and $\Teff$ is made to avoid the stars that are going through Li depletion due to convective-driven and thermohaline mixing. 
The selected samples are shown in the left panel of Fig.~\ref{Kiel_GaiaESO}. 
Note that we already excluded Li-rich stars in our samples in Fig.~\ref{Kiel_GaiaESO}, based on the definition found in \citet{2021A&A...651A..84M}, namely, A(Li)$\geq 2.0$ dex, $3800 \leq T_\mathrm{eff}/\mathrm{K}\leq 5000$, $\log g \leq 3.5$, and the gravity index $\gamma\geq 0.98$. Furthermore, the mass of each star is reported in the Gaia-ESO catalogue \citep[see][]{2021A&A...651A..84M}. We show in the right panel of Fig.~\ref{Kiel_GaiaESO} the mass distribution of stars in the two subsamples. In particular, the metal-poor stars ([Fe/H]$\leq -0.8$) are, in fact, low-mass stars with masses of approximately $ 0.8\Msun$. Within the uncertainty, they appear to be consistently below $1 \Msun$. This motivates us to explore the correction from available stellar models on the variation of Li abundances, which will be described in the next section.
We also adopted the catalogue from \citet{2021A&A...653A..72R} that includes 26 carefully selected open clusters, and  $3\,210$ field stars from the Gaia-ESO iDR6 survey. This sample also contains important information, such as the stellar ages and galactocentric distances.
We applied the same selection method as the previous sample, meaning $\log g>3.6$ and $T_\mathrm{eff}/K \geq 5850$.  

\begin{figure}[t]
    \centering
    \includegraphics[width=\columnwidth]{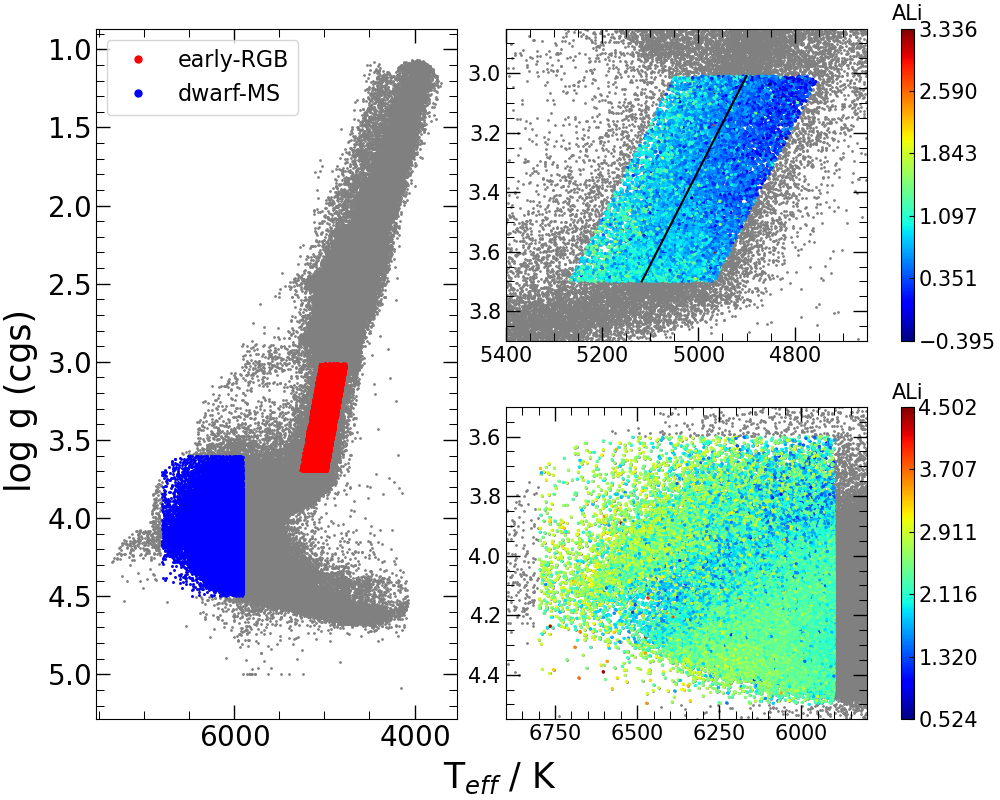}
    \caption{Left panel: the complete sample of Li detection adopted from the GALAH DR4 catalogue (grey dots). Our two selected samples are highlighted in blue and red. Top-right panel: zoom in on our selected early RGB stars sample, with the colour bar indicating A(Li). Bottom-right panel: our selected dwarf-MS stars sample. See the text for details of the selections.}
    \label{Kiel_subplots_GALAH}
\end{figure}

The GALAH survey also provides us with another opportunity to study the chemical history of the Milky Way using high-resolution spectroscopy. Li is one of the main focuses of the survey, \citet{2024MNRAS.528.5394W} provided 3D non-LTE (3D-NLTE) Li abundances for 581\,149 field stars released in GALAH DR3 \citep[][]{2021MNRAS.506..150B}. In this work, we adopted the catalogue with three additional conditions for our selected samples. They are with flags: i) \texttt{flag\_ALi} = 0, which means only 228\,613 field stars with actual Li detection are used; ii) \texttt{flag\_fe\_h} = 0 and iii) \texttt{flag\_sp} = 0, which mean no problem noted from the observational analysis. The full sample is presented on the left panel of Fig.~\ref{Kiel_subplots_GALAH}. Our subsamples for the dwarf-MS and the early RGB stars are highlighted in blue and red, respectively. 
First, for the dwarf-MS sample, we selected stars with $3.6<\log g<4.5$ and $5850\leq T_\mathrm{eff}/\mathrm{K}\leq 6800$. These choices of $\log g$ and $\Teff$ are also to avoid the contamination from stars that are undergoing the convection-driven Li depletion. We obtained about 97\,193 stars for this subsample, which are shown in the bottom-right panel of Fig.~\ref{Kiel_subplots_GALAH}. 
Then, for our early RGB stars, to avoid distraction from the red clump stars, we selected stars with $3.0<\log g<3.7$. After that, within the narrow region $4500<T_\mathrm{eff}<5500$ K, we searched for the synthetic line that mimics the RGB evolution in this region by using the simple \texttt{polyfit} python function, shown as the black line on the top-right panel of Fig.~\ref{Kiel_subplots_GALAH}. We then selected the stars within the range of $T_\mathrm{eff}^{synthetic}\pm 150$ K for our early RGB sample. As a result, we obtained about 15\,316 RGB stars for the sample.

Furthermore, we adopted the catalogue from \citet{2018A&A...615A.151B}, which contains in total 714 F- and G-dwarf, turn-off, and subgiant stars. Their Li abundances were derived with 1D-LTE analysis of high-resolution spectra. For our sample, we selected only 275 stars with the actual detected Li abundances and $\Teff \geq 5850$ K.

\section{Predictions from stellar models}\label{stellar_model}
\citet{2025A&A...696A.136N} introduced stellar models with different values of envelope overshoot efficiency parameter between the pre-main-sequence and the post-main-sequence evolutions to reproduce the Li plateau on the MS and the RGB bump location of the GC NGC~6397, using the \texttt{PARSEC} code \citep[see also][for more information]{2012MNRAS.427..127B,2022A&A...665A.126N,2025arXiv250802393N}. In particular, a value of $\Lambda_\mathrm{e}=0.6 H_p$ is applied to the post-MS evolution to calibrate the RGB bump, and a smaller value, $\Lambda_\mathrm{e}^\mathrm{p}=0.05-0.6 H_p$ (dependent on the initial masses), is applied to the early PMS evolution, where $H_p$ is the pressure scale height. 
As a result, a depletion from the primordial value ($\mathrm{A(Li)}=2.69$ dex; \citet{2014JCAP...10..050C}, adopted as the initial value applied to the stellar models) to the measured value in dwarf, turn off, and subgiant stars of NGC 6397 ($\mathrm{A(Li)}\approx 2.22$ dex; \citet{2009A&A...503..545L}) is obtained. 

\begin{figure}[t]
    \centering
    \includegraphics[width=\columnwidth]{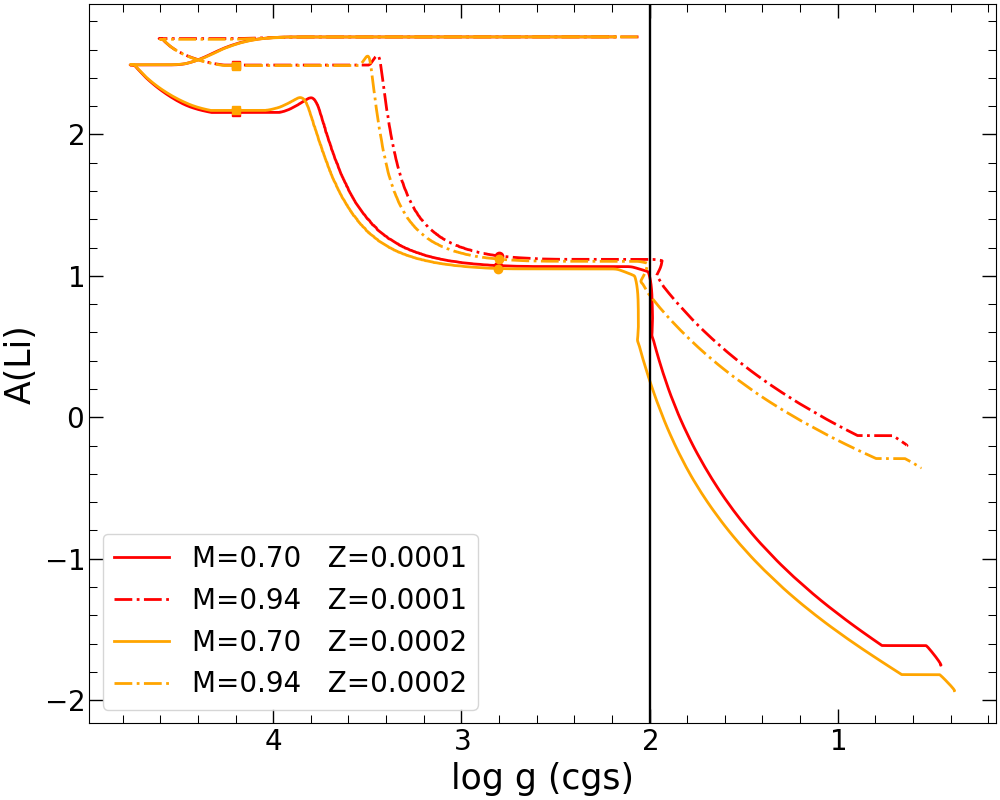}
    \caption{The variation of surface Li abundance with $\log g$ of four selected stellar models. The square symbol marks the location where $\log g\sim 4.1$ on the MS, while the circle symbol indicates $\log g \sim 2.8$ on the early RGB. The black line indicates $\log g\sim 2$ around the RGB bump. }
    \label{Ali_logg}
\end{figure}

\begin{figure}[t]
    \centering
    \includegraphics[width=\columnwidth]{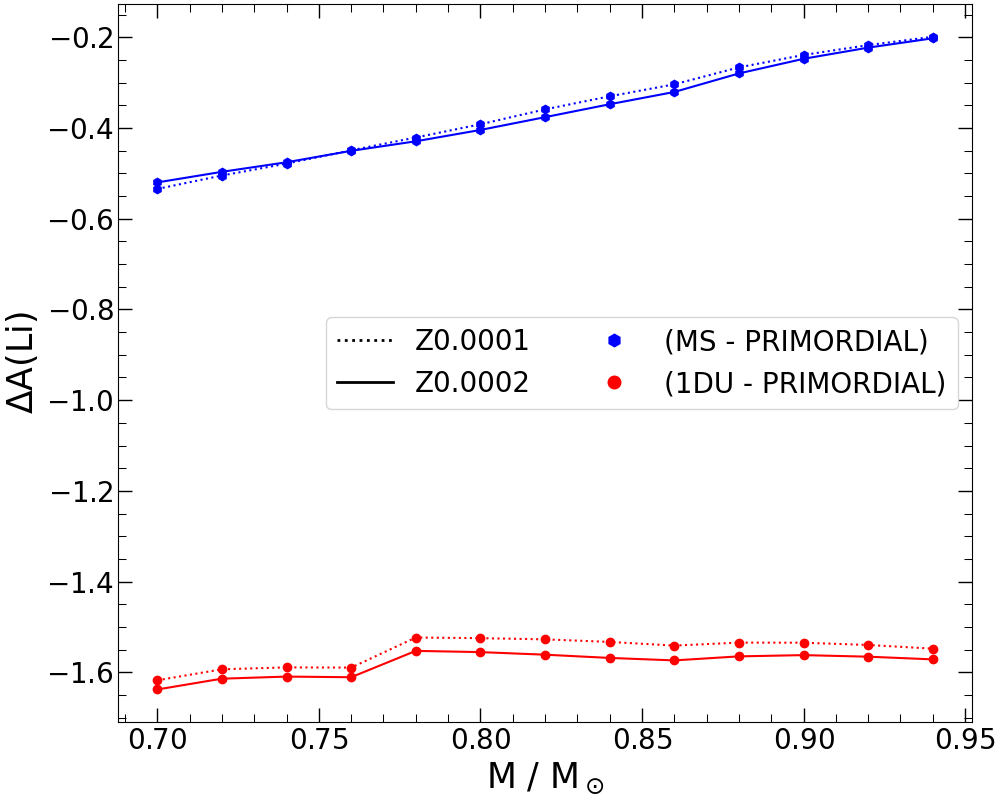}
    \caption{Depletion of Li abundance up to the MS (blue) and after the 1DU at the early RGB phase (red). The models are taken from \citet{2025A&A...696A.136N}, including the depletion during the PMS phase due to the envelope overshoot.}
    \label{DU_effect}
\end{figure}

As mentioned above, the observed metal-poor stars ($\mathrm{[Fe/H]} \leq -0.8$) show a mass range of approximately $0.8-1.0\Msun$. Meanwhile, \citet{2025A&A...696A.136N} provides models with $M\sim 0.7-0.94\Msun$ for two metallicity sets $Z=0.0001,0.0002$ ($\mathrm{[Fe/H]}\approx -2.4,-2.1$). We show the variation of $\mathrm{A(Li)}$ with $\log g$ for four selected models at the limit of the sets in Fig.~\ref{Ali_logg}. 
All models evolve from the PMS (starting at $\log g\sim 2.2$, $\mathrm{A(Li)}=2.69$) towards higher $\log g$ and make the turning point at the zero-age-MS (ZAMS). The variation of $\mathrm{A(Li)}$ during this phase shows a clear depletion due to the efficiency of the envelope overshoot, which mainly depends on mass. Subsequently, atomic diffusion causes further depletion of A(Li) from the ZAMS until the end of the MS, where mixing partially restores it. 
The squared symbols mark the location where diffusion shows its maximum efficiency. 
Subsequently, the convective envelope penetrates toward inner layers, 
which leads to the dilution of Li in the envelope. This dilution process leads to a significant depletion during this evolutionary phase until it reaches a constant value when the first dredge-up (1DU) is completed in the RGB phase (dotted symbols). 
Thermohaline mixing becomes active and depletes even more A(Li) in the more advanced evolution above the RGB bump \citep[see also][and references therein]{2007A&A...467L..15C}. 

In Fig.~\ref{DU_effect}, we show the depletion of Li abundance from the initial value ($\mathrm{A(Li)}=2.69$ dex) at the MS ($\log g\sim 4.1$) and at the RGB (before the so-called RGB bump) when the 1DU is already completed ($\log g\sim 2.8$). The theoretical prediction indicates an overall depletion from the initial value throughout the entire evolution of the stars. Firstly, within the presented mass and metallicity ranges, a depletion of $\sim 0.2-0.5$ dex is seen at the MS phase, starting from the primordial value. Especially in the mass range $\leq 0.8\Msun$, model prediction shows a depletion of $\sim 0.4-0.5$ dex, which is the typical value that is needed to reproduce the Spite plateau \citep[$\mathrm{A(Li)}\approx 2.2$ dex;][]{1982A&A...115..357S,1982Natur.297..483S} from the cosmologically determined value. 

Secondly, within the typical uncertainty ($\approx 0.1$ dex), the depletion after the 1DU from the initial value shows an approximately constant value, that is, $\Delta \mathrm{A(Li)} \approx 1.6$ dex for all shown models. This prediction might indicate the existence of a second plateau that is located at the early RGB evolution, where the stars have not yet been affected by thermohaline mixing, which confirms the observational results obtained by \citet{2022A&A...661A.153M}. 
This aspect will be discussed in the results section below. 

\section{Chemical evolution model}\label{cem}

\subsection{Model for thin disc}
In order to study the Li evolution of the thin disc of the Milky Way, we adopted the model introduced by \citet{2019MNRAS.482.4372C}. We point the reader to \citet{2019MNRAS.482.4372C} for more details about the model, while we here only summarise the main ingredients for our work. 
In particular, the initial mass function (IMF) is adopted from \citet{2001MNRAS.322..231K}, and the stellar lifetime follows \citet{2002A&A...390..561M}. The single degenerate scheme for the progenitors of SNe Ia is adopted from \citet{1986A&A...154..279M}. 
We assume the thin disc to be formed by slow gas infall (for details see \citet{2019MNRAS.482.4372C}).

The key ingredient of this model for Li is the assumption that nova systems are the principal producers of Li, although we also consider a small contribution from cosmic rays 
\citep[see][]{2012A&A...542A..67P}. 
The nova rate is computed by assuming the same delay time distribution function (DTD) of the SNe Ia deriving from cataclysmic variables (single degenerate scenario) as in \citet{1986A&A...154..279M}, but with an additional time delay to allow for the cooling of the white dwarfs.

In \citet{2019MNRAS.482.4372C}, it was assumed that only binary systems formed by stars in the mass range $M=0.8-8\Msun$ can develop nova systems. 
The probability of forming a binary system of a certain mass is weighted on the IMF. The configurations of primary and secondary stars are defined by
\begin{align}
    f(\mu) = 2^{1+\gamma}(1+\gamma)\mu^\gamma.
\end{align}
In which, $f(\mu)$ is the distribution function for the mass fraction of the secondary star, $\mu=M_2/M_\mathrm{B}$, where $M_2$ is the mass of the secondary star and $M_\mathrm{B}$ is the mass of the binary system. The exponential coefficient $\gamma = 2$ is adopted from \citet{1983A&A...118..217G}. 

The fraction of binary systems that develop a nova thus becomes a crucial parameter in the model. This parameter is defined by $N_\mathrm{lithium}$. In this work, we keep using the proposed fraction by \citet{2019MNRAS.482.4372C}, that is $N_\mathrm{lithium} = 0.03$, which corresponds to a Galactic rate of nova bursts at present of $\sim 20-30$ yr$^{-1}$ \citep[see][]{1997ApJ...487..226S,2020A&ARv..28....3D}. 

The time when Li production takes place is defined as the time at which the primary star evolves into a white dwarf plus a so-called delay time, $\tau_\mathrm{nova}$. In particular, $\tau_\mathrm{nova}$ is the time needed for a white dwarf to ignite the first nova outburst, which can occur only after the cooling of the white dwarf.  \citet{2019MNRAS.482.4372C} found $\tau_\mathrm{nova}=1$ Gyr to be the best constrained value, and in agreement with the previous works \citep[e.g.][]{2001A&A...374..646R,2015ApJ...808L..14I}, thus we adopted this value in this work. 

The total Li produced by a nova during its lifetime is defined as $^{Li}Y_\mathrm{Nova}$. The assumption that all novae produce the same amount of Li in all events is also adopted in this work. 
A value of $^{Li}Y_\mathrm{Nova}=1.8\times 10^{-5} \Msun$, with a typical number of $10^4$ outbursts events during the entire lifetime of novae \citep[][]{1978ApJ...219..595F}, is also suggested  in the model of \citet{2019MNRAS.482.4372C}, where they assumed the measured Li abundance from the dwarf halo stars, $\mathrm{A(Li)}=2.28$ dex, as the initial gas composition.  For many years, this abundance has been assumed to be the primordial one until the Wilkinson Microwave Anisotropy Probe (WMAP) and Planck satellites suggested that the primordial  Li abundance is three times higher \citep[$\mathrm{A(Li)}=2.69$ dex, e.g.][]{2003PhLB..567..227C,2004ApJ...600..544C}.

\begin{figure}[t]
    \centering
    \includegraphics[width=\columnwidth]{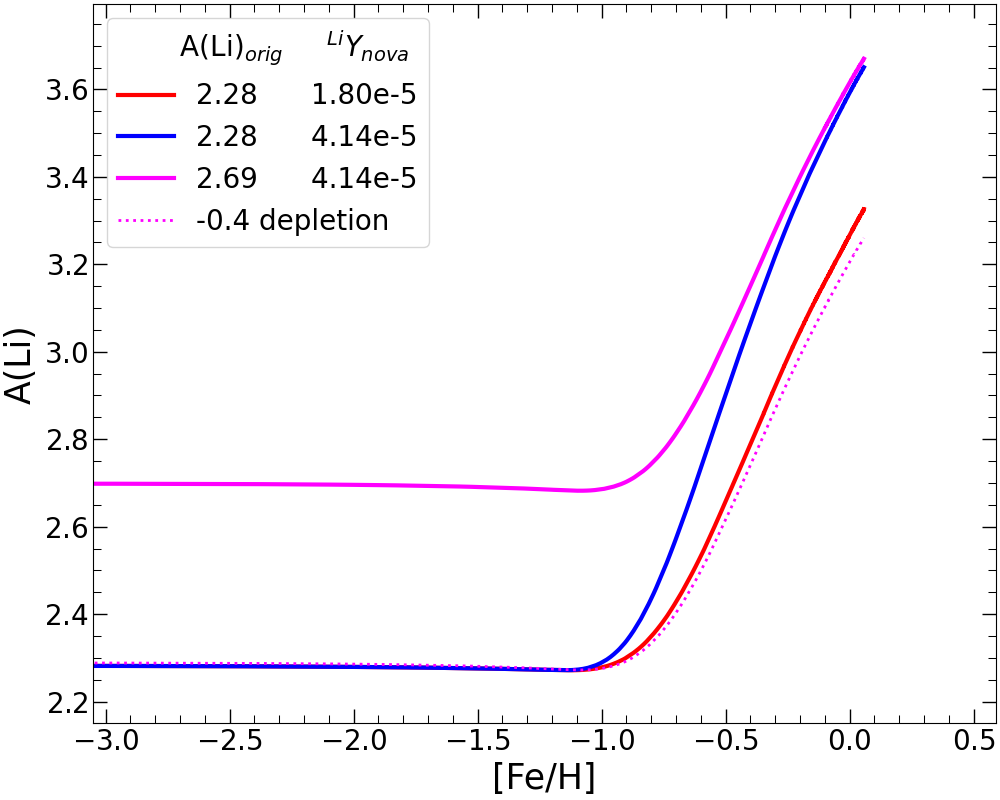}
    \caption{Chemical evolution models of Li with different adopted values of original Li abundance and novae yields, as listed in the figure legend. An illustration model with $-0.4$ dex depletion from the primordial $\mathrm{A(Li)}=2.69$ model (magenta line) is plotted as the dotted line. }
    \label{CEM}
\end{figure}

The predicted evolution of $\mathrm{A(Li)}$ with metallicity $\mathrm{[Fe/H]}$ is shown in Fig.~\ref{CEM}.  It is clear that since Li is mainly produced by novae on long timescales, the abundance of Li for $\mathrm{[Fe/H]}<-1$ is the initial one, which could be adopted as either the primordial value or the measured Spite plateau value. On the other hand, the contribution from nova explosions is dominant in the more metal-rich region. It is worth noting that the chemical evolution model predicts the evolution of Li in the interstellar medium, and when we compare our results with the Li abundance measured in the atmospheres of stars, we assume that it has not been depleted, and therefore, we aim to reproduce the upper envelope of the data. However, if the Li has indeed been depleted, this comparison should take the depletion into account. In the figure, we show models starting from an initial Li abundance as observed in halo stars ($2.28$ dex) and models starting with the primordial Li suggested by Planck ($2.69$ dex). The comparison between models with the same $^{Li}Y_\mathrm{Nova}$ but different original $\mathrm{A(Li)}$ values shows a clear trend of convergence towards higher metallicity. This means that the production of novae overcomes the difference in original abundances as the galaxy evolves. However, if a model with the primordial Li as its original composition, and a depletion due to the evolution of stars is assumed (dotted line), a higher nova Li yield is required to reproduce the prediction of the model with a lower original $\mathrm{A(Li)}$. This was suggested in \citet{2019MNRAS.482.4372C}, where $^{Li}Y_\mathrm{Nova}=4.14\times 10^{-5} \Msun$ was obtained by using the primordial $\mathrm{A(Li)}$ as the original abundance and assuming a fixed amount of depletion. 

In this work, we explore the model computed with the primordial Li abundance, $\mathrm{A(Li)}=2.69$ dex, and apply the depletion corrections from stellar models provided in \citet{2025A&A...696A.136N}. In other words, the evolution of Li is expressed as
\begin{align}
    A(\mathrm{Li}) = A(\mathrm{Li})_\mathrm{CEM} + \Delta A(\mathrm{Li})_\mathrm{SM},
\end{align}
where $A(\mathrm{Li})_\mathrm{CEM}$ is obtained from the chemical evolution model, with the variables are the original Li abundance and the nova Li yield that applied to the model. $\Delta A(\mathrm{Li})_\mathrm{SM}$ is the predicted correction from stellar models, which depends on mass and metallicity, as will be shown below. 

\subsection{Correction from stellar models}\label{stellar_correction}
As described in Sect.~\ref{stellar_model}, the grid of stellar models covers the mass ranges from $0.70-0.94\Msun$ and metallicity $-2.4\leq \mathrm{[Fe/H]}\leq -2.1$ are taken from \citet{2025A&A...696A.136N}. Within the models, a depletion $\Delta A(\mathrm{Li})\approx 0.2-0.5$ dex from the primordial value ($\mathrm{A(Li)}=2.69$ dex) at the MS phase mainly depends on the stellar masses. On the other hand, a depletion of around $1.52 - 1.63$ dex is predicted for stars at the early RGB phase, depending on mass and metallicity. 
Therefore, to determine the amount of depletion at a given point of the Li evolution (in the plane of $\mathrm{A(Li)}$ versus $\mathrm{[Fe/H]}$), the metallicity and mass at this point are needed. 

\begin{figure}[t]
    \centering
    \includegraphics[width=\columnwidth]{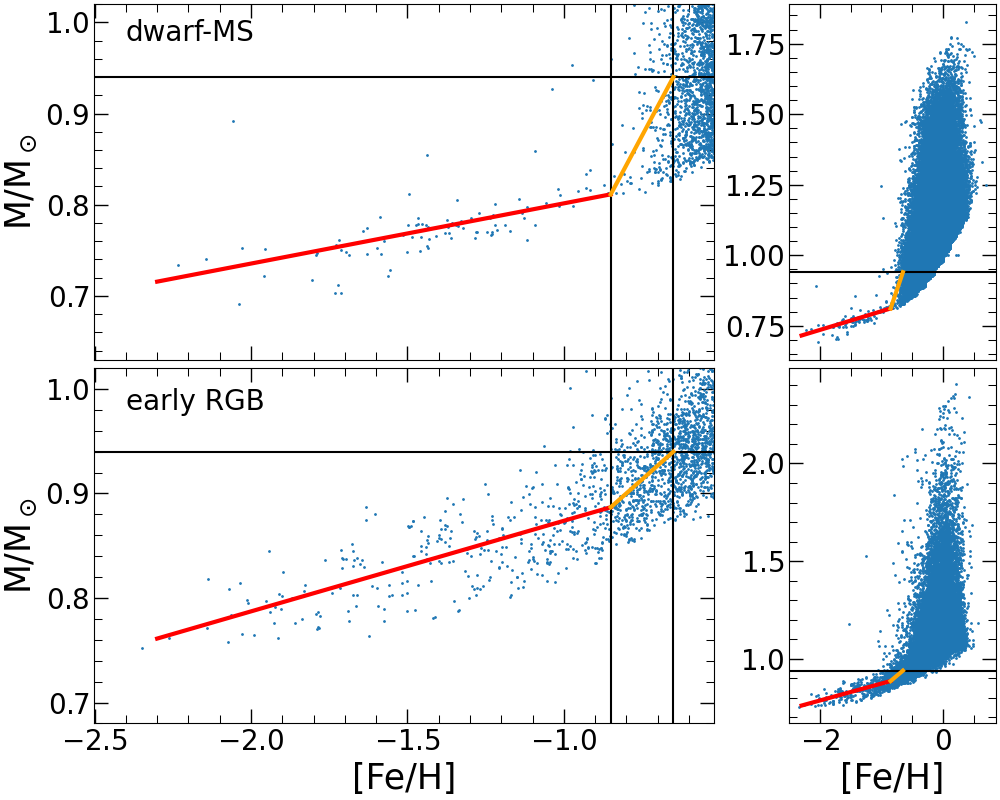}
    \caption{Mass distribution with metallicity in two samples of dwarf-MS (top panel) and early RGB (bottom panel) selected from the GALAH catalogue. The left panels zoom in on the low-mass and metal-poor area. The right panels show the whole sample. The empirical relations at two regions of [Fe/H] are shown in red and orange solid lines. The black horizontal line indicates the mass limit $0.94\Msun$, while the vertical lines indicate the metallicity $\mathrm{[Fe/H]}=-0.85$ and $-0.65$. }
    \label{galah_mass_FeH}
\end{figure}

In Fig.~\ref{galah_mass_FeH}, we show the mass distribution of the selected dwarf-MS and early RGB stars from the GALAH catalogue (top- and bottom-panels correspondingly). The figure indicates the lowest mass presented in the samples is $\sim 0.7\Msun$, which is covered by the stellar grid of \citet{2025A&A...696A.136N}. It also clearly indicates different trends between stars below and above $\mathrm{[Fe/H]}\sim -0.75$, especially in the case of dwarf-MS stars. However, the stellar grid that is used in this work is limited to $0.94\Msun$. Therefore, we adopt the depletion predicted by this $0.94\Msun$ model for higher masses. 
We divide the samples into three regions: below $\mathrm{[Fe/H]}\leq -0.85$; between $-0.85< \mathrm{[Fe/H]}< -0.65$; and above $\mathrm{[Fe/H]}\geq -0.65$. We used the \texttt{polyfit} function to obtain the empirical relation of mass and metallicity for the stars within the first two regions, which are plotted in red and orange in Fig.~\ref{galah_mass_FeH}. As a result, we obtained the relations,
\begin{align}
    & M = 0.066\mathrm{[Fe/H]} + 0.867, \mathrm{\ for\ \mathrm{[Fe/H]}\leq -0.85,}\\
    & M = 0.643\mathrm{[Fe/H]} + 1.358, \mathrm{\ for\ \mathrm{[Fe/H]}= [-0.85,-0.65]}.
\end{align}
Similarly, we obtained the relations for early RGB stars as follows,
\begin{align}
    & M = 0.086\mathrm{[Fe/H]} + 0.960, \mathrm{\ for\ \mathrm{[Fe/H]}\leq -0.85,}\\
    & M = 0.265\mathrm{[Fe/H]} + 1.112, \mathrm{\ for\ \mathrm{[Fe/H]}= [-0.85,-0.65]}.
\end{align}

The empirical relations above establish the relation between mass and metallicity. In other words, at a given metallicity $\mathrm{[Fe/H]}$ along the evolution of Li, we can deduce the mass and thus determine the amount of depletion that is predicted from the stellar grid. For this purpose, we applied the linear interpolation method in mass and metallicity from the stellar grid to obtain the predicted amount of depletion at a given point of the galactic $\mathrm{A(Li)}$ evolution. In this regard, it should be noted that, for masses below $0.7\Msun$, the predicted depletion of the $0.7\Msun$ model is adopted. For masses above $0.94\Msun$, the predicted depletion of the $0.94\Msun$ model is adopted. 

\begin{table}[!tbp]
\caption{Adopted parameters in each model.
}
\label{models_table}
\centering
\begin{tabular}{l | c c c}
\hline\hline
Model & $\mathrm{A(Li)_{orig}}$ & $^{Li}Y_\mathrm{Nova}$ & depletion \\
\hline
model C2.34a & $2.69$ & $2.34\times 10^{-5}\Msun$ & no \\
model S1.80a & $2.28$ & $1.8\times 10^{-5}\Msun$ & no \\
model C2.34b & $2.69$ & $2.34\times 10^{-5}\Msun$ & yes, MS \\
model C2.02b & $2.69$ & $2.02\times 10^{-5}\Msun$ & yes, MS \\
model C2.34c & $2.69$ & $2.34\times 10^{-5}\Msun$ & yes, RGB \\
model S1.80c & $2.28$ & $1.8\times 10^{-5}\Msun$ & yes, RGB \\
model C2.02c & $2.69$ & $2.02\times 10^{-5}\Msun$ & yes, RGB \\
\hline
\end{tabular}
\tablefoot{
Models are named according to their adopted parameters, following the format $XY.YYz$, where $X$ represents the initial $\mathrm{A(Li)}$ - either the Cosmological (C) or Spite plateau (S) values; $Y.YY$ is $^{Li}Y_\mathrm{Nova}/10^{-5}$ value; and $z$ is the letter denoting the stellar Li depletion.
}
\end{table}

\section{Results}\label{results}
As mentioned above, the models presented in this work adopt the primordial Li, which is constrained by the number of baryons per photon from the WMAP satellite or the cosmological parameters determined by Planck \citep[][]{2011ApJS..192...18K,2014JCAP...10..050C}, that is $\mathrm{A(Li)}=2.69$ dex. 
The models, with or without the Li depletion corrections predicted from stellar models (Sect.~\ref{stellar_model}), are then directly compared to galactic field stars at two evolutionary phases, namely, the dwarf-MS and early RGB (below the RGB bump), as selected in Sect.~\ref{sample_selection}. 
Different models with varying values of original $\mathrm{A(Li)}$ and $^{Li}Y_\mathrm{Nova}$ are summarised in Table~\ref{models_table}.

\subsection{Dwarf main sequence stars}

\begin{figure}[t]
    \centering
    \includegraphics[width=\columnwidth]{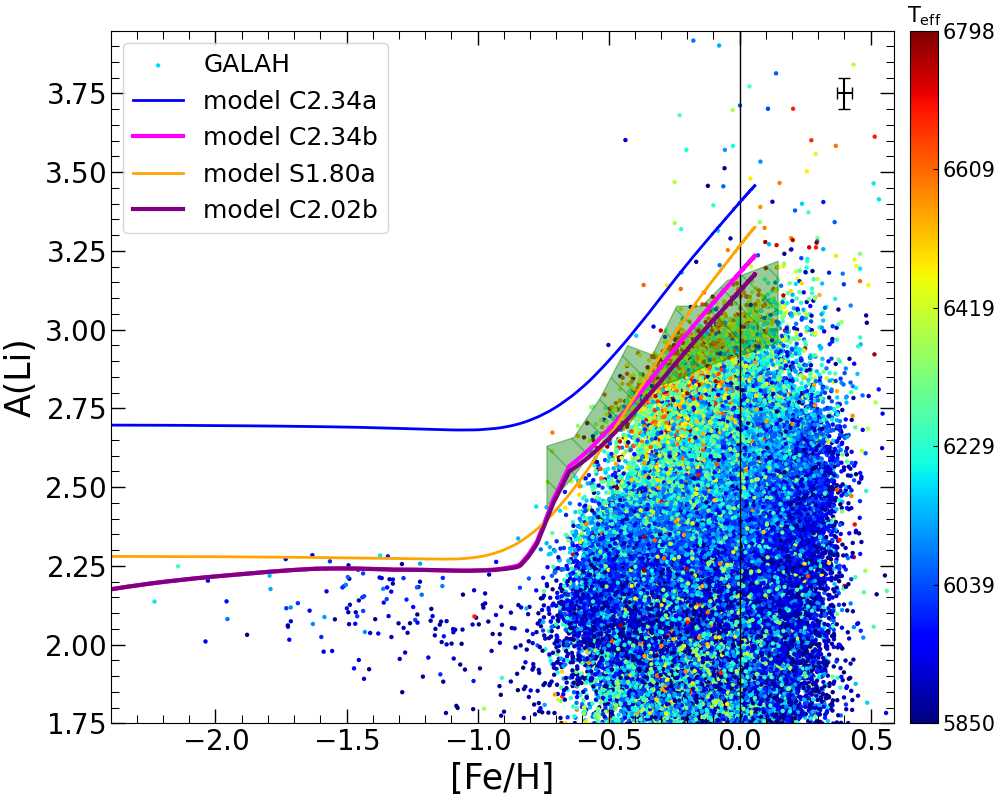}
    \caption{Lithium abundances versus metallicity of dwarf-MS stars from GALAH sample, colour-coded by their effective temperatures. Our models with and without stellar correction are shown in magenta and blue lines, respectively (see Table.~\ref{models_table}). The defined upper envelope is shown by the green-shaded area. The error bars indicate the mean uncertainties of the observed data. The black vertical line marks for $\mathrm{[Fe/H]}=0$.}
    \label{galah_Ali_FeH_MS}
\end{figure}

First of all, we compute models with different nova Li yields, assuming the original Li abundance as $\mathrm{A(Li)}=2.69$ dex. The correction from stellar models is then added to the simulated Li abundance, following the description in Sect.~\ref{stellar_correction}.

Secondly, we define an upper envelope for the stars with $\mathrm{[Fe/H]}\geq -0.8$ dex by computing the mean value of $0.1\%$ highest $\mathrm{A(Li)}$ data points in each bin over 10 bins of 0.1 dex in $\mathrm{[Fe/H]}$, with the minimum number of stars set to 5. The envelope is then constructed by one standard deviation from the mean values of each bin, as shown in Fig.~\ref{galah_Ali_FeH_MS}. 
Meanwhile, the large spread of $\mathrm{A(Li)}$ in this region might have many mechanisms at work, for instance, the stellar internal rotation and angular momentum transport processes \citep[e.g.][and references therein]{2022NatAs...6..788E}, or the intrinsic properties of those stars \citep[e.g.][and references therein]{2025A&A...699A.173D}. 
Additionally, there are stars with extremely high $\mathrm{A(Li)}$ in the region with $\mathrm{[Fe/H]}> -0.4$ dex. As claimed in the literature, the origins of these stars remain unclear \citep[see][]{2012MSAIS..22...79K,2020MNRAS.497L..30G}. However, an investigation into these stars is beyond the scope of this paper and is worth a separate study. 

As a result, Fig.~\ref{galah_Ali_FeH_MS} implies model C2.34b, computed with the primordial value $\mathrm{A(Li)}=2.69$, $^{Li}Y_\mathrm{Nova}=2.34\times 10^{-5}\Msun$ and taking into account the correction from stellar model, is the best-fit model. At low metallicities ($-2.5<\mathrm{[Fe/H]}< -0.8$ dex), our model reproduces very well the Spite plateau, with $\mathrm{A(Li)}=2.15-2.25$ dex. At higher metallicities, our best-fit model reproduces very well the defined upper envelope. At $\mathrm{[Fe/H]}=0$, our model shows $\mathrm{A(Li)}=3.16$ dex and a depletion of $-0.22$ dex predicted from stellar model, which is consistent with the measured meteoritic value, $\mathrm{A(Li)}\sim 3.33$ \citep[][]{2009LanB...4B..712L}. 

The Li yield predicted by our model is in good agreement with the mean value estimated from observations of nova beryllium-7 (Be) yield, that is $\sim 2.02 \times 10^{-5} \Msun$ \citep[see][]{2023MNRAS.518.2614M,2025A&A...698A.291I}. For better comparison, we compute a model adopting this mean value, named as model C2.02b. As expected, the Spite plateau is unaffected by changes in the nova Li yield, while lowering the yield results in a lower Li abundance at the high metallicity area. The discrepancy between predictions of the two models is increasing at higher metallicities, for example, a modest difference of $\sim 0.05$ dex in $\mathrm{A(Li)}$ is seen at the solar metallicity. Regardless of the modest discrepancy, both models successfully reproduce the defined upper envelope. Furthermore, the yield obtained by our best-fit model is well covered by the observed data of nova Be yields, which span from $3.4\times 10^{-6}$ to $9.7\times 10^{-5} \Msun$ \citep[see also][]{2016MNRAS.463L.117M,2021ApJ...916...44A}. This result, obtained with our model, reinforces the role of novae production in the enrichment of galactic Li. Nonetheless, we should also stress that our chemical evolution model employs a simple scheme to take into account the contribution from nova explosions, assuming that all novae produce the same amount of Li in all events. 
The contribution from novae is more complex, and we expect it to depend on metallicity and mass, as shown in the works of \citet{2022ApJ...933L..30K} and \citet{2024A&A...689A.222K}. A more complex treatment of novae yields will most likely improve the chemical evolution model in the solar vicinity; however, it is beyond the scope of this paper and will be addressed in the forthcoming work.

\begin{figure}[t]
    \centering
    \includegraphics[width=\columnwidth]{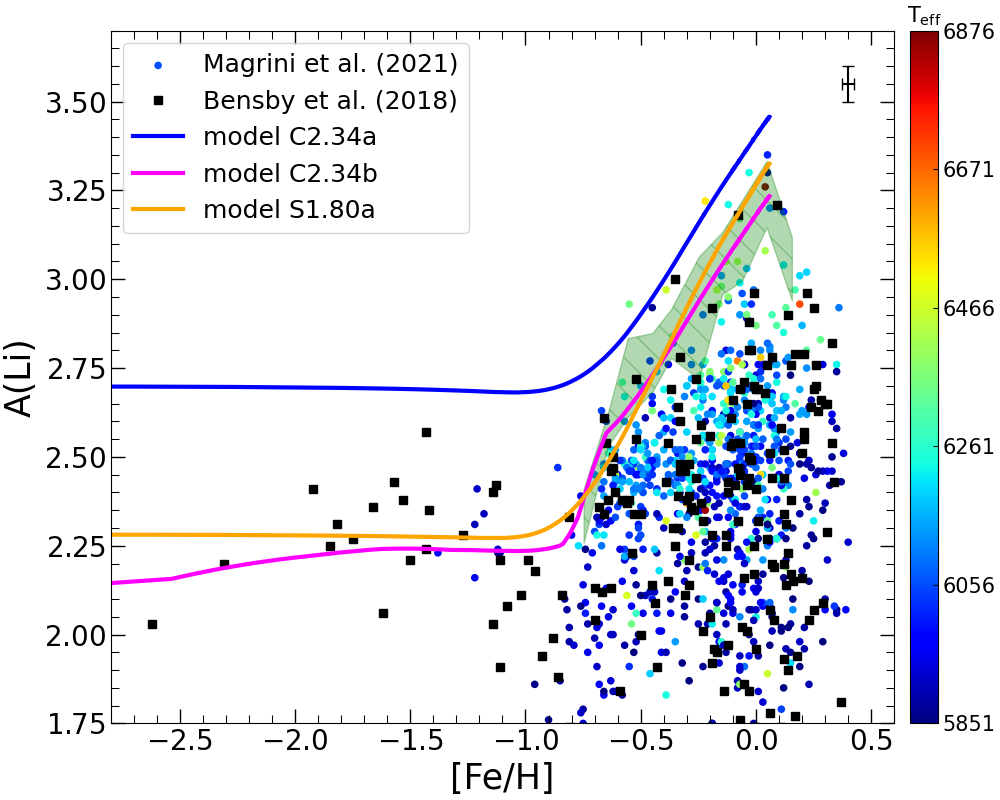}
    \caption{Li-abundance versus metallicity of dwarf-MS field stars from the Gaia-ESO \citep[][]{2021A&A...651A..84M} and Bensby et al. catalogues. The models shown in Fig.~\ref{galah_Ali_FeH_MS} are also adopted in this plot. The black error bars indicate the mean uncertainties of the Bensby et al. sample.}
    \label{GaiaESO_Ali_FeH_MS}
\end{figure}

The sample of dwarf-MS field stars from the Gaia-ESO catalogue is shown in Fig.~\ref{GaiaESO_Ali_FeH_MS}. We also included 275 stars from \citet{2018A&A...615A.151B} with the metallicity extending to $\sim -2.6$ dex. 
The upper envelope is drawn from the sample of Gaia-ESO data \citep{2021A&A...651A..84M} for the comparison. Within the typical uncertainty of A(Li) that is $\sim 0.1$ dex, our model reproduces very well the Spite plateau presented in this sample. The comparison to the defined upper envelope, drawn from the Gaia-ESO data, suggests the model with $^{Li}Y_\mathrm{Nova}=2.34\times 10^{-5}\Msun$ is the best-fit model. This obtained result is in complete agreement with the GALAH data presented in Fig.~\ref{galah_Ali_FeH_MS}. 

\begin{figure}[t]
    \centering
    \includegraphics[width=\columnwidth]{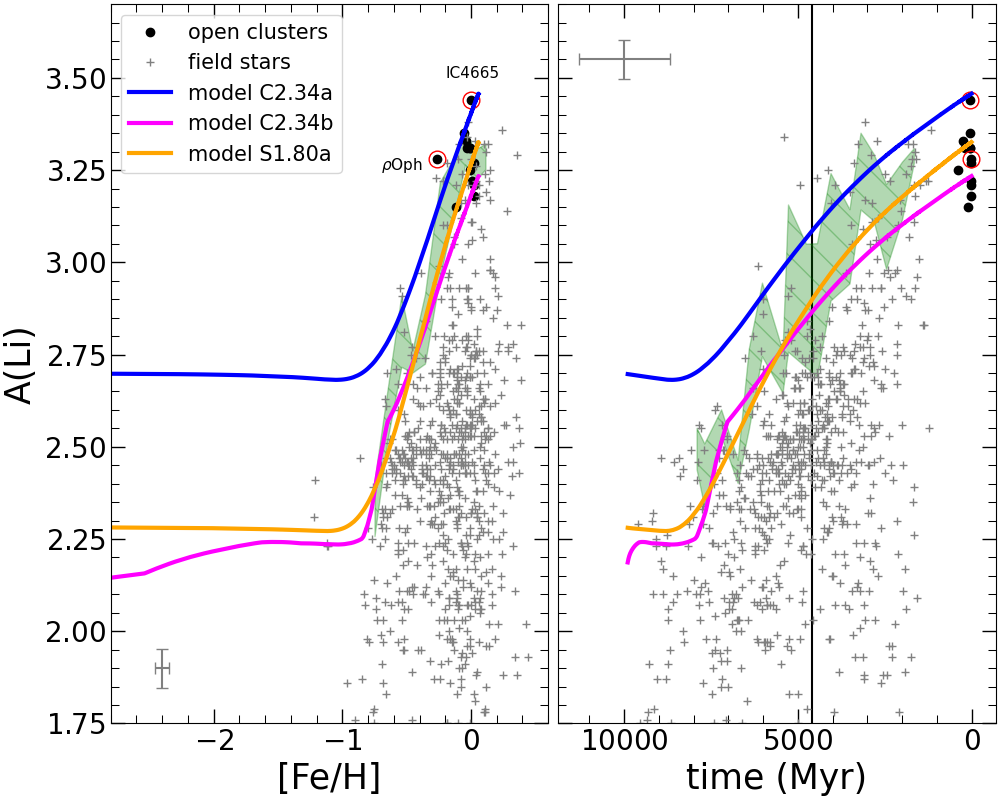}
    \caption{Li abundances of open clusters and field stars from the Gaia-ESO catalogue \citep{2021A&A...653A..72R} as a function of metallicity (left panel), and stellar age (right panel). The models shown in Fig.~\ref{galah_Ali_FeH_MS} are also adopted in this plot. 
    The grey error bars indicate the mean uncertainties. The shaded area indicates the defined envelope, computed from the field stars. The black vertical line on the right panel is for an age of $4.6$ Gyr.}
    \label{GaiaESO_Ali_age_MS}
\end{figure}

The sample of $\sim 700$ field stars and the averaged Li abundances of 13 open clusters in the solar vicinity (with the galactocentric distance between $7-9$ kpc), selected from the catalogue of \citet{2021A&A...653A..72R}, are shown in Fig.~\ref{GaiaESO_Ali_age_MS}. 
In the left panel, the upper envelope of the field stars in this sample shows a remarkable agreement with the other three catalogues for our best-fit model (model C2.34b). 
Most clusters have $\mathrm{A(Li)}$ within the defined envelope, except for two clusters that are circled in red. However, we also notice their large uncertainties, which are $\sigma_{\mathrm{A(Li)}}=0.1$ and $0.13$ dex for IC~4665 and $\rho$Oph, respectively. 
The right panel of Fig.~\ref{GaiaESO_Ali_age_MS} shows Li abundances as a function of age (in Myr). 
In this case, we define the upper envelope by considering 20 bins of $500$ Myr, starting from $1000$ Myr. 
The black solid line indicates the time when the Sun formed, 4.6 billion years ago. The prediction of the model for the interstellar medium at this age is $\mathrm{A(Li)} \sim 3.1$ dex, which is $0.2$ dex lower than the meteoritic value \citep[$\mathrm{A(Li)}\sim 3.3$, ][]{2009LanB...4B..712L}. This may indicate that the Sun was born from an inner galactic region, compared to stars that are present now in the solar vicinity. However, we should clarify two points: first, the uncertainties in the stellar ages prevent us from drawing firm conclusions based on this diagram; second, we consider the yields from novae and their formation channels to be constant. The possible metal and mass dependencies could influence and improve the fit as mentioned above. We decide to reserve a more complex treatment for future work. 
In any case, we find that our model successfully reproduces the Spite plateau and the Li-enrichment in the solar vicinity, showing consistency among four separate catalogues.

In addition, we show in Figs.~\ref{galah_Ali_FeH_MS}, \ref{GaiaESO_Ali_FeH_MS} and \ref{GaiaESO_Ali_age_MS} the model computed with the Spite plateau value ($\mathrm{A(Li)}=2.28$ dex) for the sake of comparison. We find agreement with the previous work \citep[][]{2019MNRAS.482.4372C}, indicating that a Li yield of $^{Li}Y_\mathrm{Nova}=1.8\times 10^{-5}\Msun$ would be the best constrained value if the original gas composition is assumed to be the measured value of Spite plateau. On the other hand, this model predicts a Li abundance in the presolar material that is lower than the measured value \citep[see the discussion of][]{2019MNRAS.482.4372C}. 

\subsection{Early RGB stars}

\begin{figure}[t]
    \centering
    \includegraphics[width=\columnwidth]{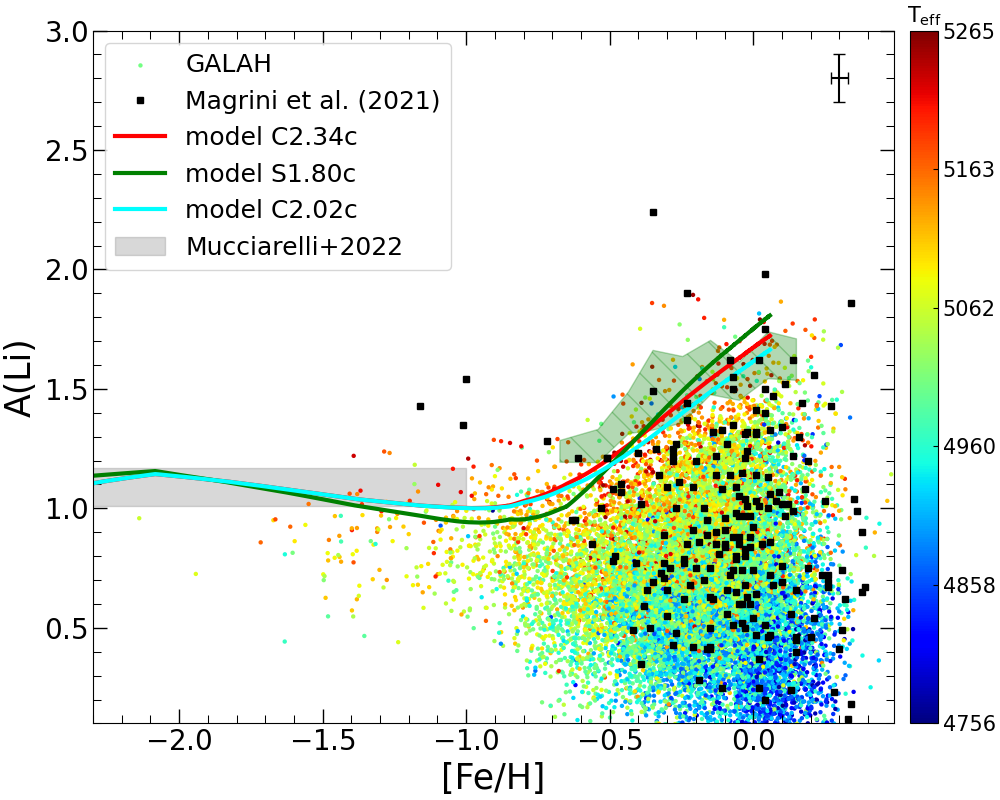}
    \caption{Li abundance versus metallicity of early RGB stars from the GALAH and Gaia-ESO catalogues, superimposed with our models (solid lines). The grey-shaded area is the early RGB plateau from \citet{2022A&A...661A.153M} ($\mathrm{A(Li)}=1.09\pm 0.08$). The black error bars indicate the mean uncertainties from GALAH's sample.}
    \label{Ali_FeH_RGB}
\end{figure}

In Sect.~\ref{stellar_model}, we found that stellar models with different masses and metallicities show a possible plateau at the RGB evolution after the 1DU and before the RGB bump. The depletion from the primordial value ($\mathrm{A(Li)}=2.69$ dex) is $\sim 1.6$ dex at this evolutionary stage. 

In Fig.~\ref{Ali_FeH_RGB}, we explore this finding with a sample selected from the GALAH catalogue. Our selected sample was described in Sect.~\ref{sample_selection}. Moreover, for the fitting purposes, we excluded the Li-rich stars by using the definition of \citet{2021A&A...651A..84M}. In total, we obtained a sample of 15\,297 early RGB stars from the GALAH catalogue, with $-2.14\leq \mathrm{[Fe/H]}\leq 0.48$, $4750\leq T_\mathrm{eff}/\mathrm{K}\leq 5270$, and $3\leq \log g\leq 3.7$. 
Indeed, the data shown in Fig.~\ref{Ali_FeH_RGB} indicate a plateau feature of $\mathrm{A(Li)}$ in the low metallicity region ($\mathrm{[Fe/H]}\leq -0.9$). At the upper limit, $\mathrm{A(Li)}$ is found to be in the range of $\sim 1.0-1.2$ dex. Meanwhile, $\mathrm{A(Li)}$ is about $1.8-1.9$ dex at $\mathrm{[Fe/H]}=0$. These typical values are about $1.1-1.3$ dex lower than those of the dwarf-MS stars in Figs.~\ref{galah_Ali_FeH_MS} and \ref{GaiaESO_Ali_FeH_MS}. 

Our chemical evolution model with the depletion from stellar evolution correction (Model C2.34c), assuming $\mathrm{A(Li)}=2.69$, reproduces very well the early RGB plateau, as indicated in Fig.~\ref{Ali_FeH_RGB}. Moreover, this model, computed with $^{Li}Y_\mathrm{Nova}=2.34\times 10^{-5}\Msun$, also reproduces very well the upper envelope at the Li-enrichment area. 
The chemical evolution model, computed with $\mathrm{A(Li)}=2.28$ dex as the original Li abundance and taking into account the Li depletion between the MS and early RGB phases, is plotted as the solid green line in Fig.~\ref{Ali_FeH_RGB} and is named model S1.80c. The prediction by this model is very similar to that of model C2.34c. This result, on one hand, prevents us from drawing a firm conclusion about the original Li abundance. On the other hand, it also implies the crucial role of stellar Li depletion correction when studying the galactic Li evolution by using these early RGB stars.

Overall, our chemical evolution model with stellar evolution correction for dwarf-MS and early RGB, assuming the primordial Li abundance, achieves a significant agreement with observations. 
In particular, in the metal-poor region, by taking into account the Li depletion from stellar models, our chemical evolution model (model C2.34b) reproduces very well the Spite plateau, $\mathrm{A(Li)}\sim 2.2$ dex. 
At the same time, model C2.34c reproduces very well the plateau of early RGB stars found by \citet{2022A&A...661A.153M}, with an averaged $\mathrm{A(Li)}\sim 1.1$~dex. 
In the high metallicity region, where nova explosions sensibly contribute to the production of Li, our models C2.34b and C2.34c, with $^{Li}Y_\mathrm{Nova}=2.34\times 10^{-5}\Msun$, reproduce very well the enrichment of galactic Li data in both the early RGB and dwarf-MS stars. 

We also include the early RGB stars from the catalogue of \citet{2021A&A...651A..84M} in Fig.~\ref{Ali_FeH_RGB}. At low metallicities, no $\mathrm{A(Li)}$ plateau is found in their data. This can be understood from the fact that their catalogue is limited to $\mathrm{[Fe/H]}>-1$ dex. At higher metallicities, our model with stellar Li depletion provides a satisfactory fit to the data. 

Finally, we should emphasise that this analysis of the early RGB stars is, in fact, the first attempt to study the behaviour of Li due to the 1DU and earlier mixing processes (e.g. convective-driven and atomic diffusion). We rely on the defined upper envelope to obtain our results, as usually done in the literature \citep[][]{1988A&A...192..192R,1999A&A...352..117R,2019MNRAS.482.4372C,2021A&ARv..29....5M}. However, the presence of Li-rich stars at this evolutionary stage can be problematic for a precise definition of the upper limit \citep[see][and references therein]{2016MNRAS.461.3336C,2021MNRAS.505.5340M}. A better understanding of Li-rich stars is reserved for future work.

\section{Conclusions}\label{conclusion}
We present in this paper the chemical evolution models for Li in the thin disc of the Milky Way. We adopted data from GALAH DR4 \citep[][]{2024MNRAS.528.5394W}, the Gaia-ESO surveys \citep[][]{2021A&A...651A..84M,2021A&A...653A..72R}, and the sample of \citet{2018A&A...615A.151B} for our comparison purposes. 

We paid particular attention to two phases of stellar evolution, namely the dwarf-MS and early RGB. The data reveal the presence of $\mathrm{A(Li)}$ plateaus, namely the renowned Spite plateau of dwarf-MS stars and the early RGB plateau. The second plateau was recently discovered by \citet{2022A&A...661A.153M} and is also identified in our sample. 

The depletion of Li abundance from the primordial value ($\mathrm{A(Li)}=2.69$ dex) during the PMS evolution of low-mass metal-poor stars due to the efficiency of envelope overshoot, was recently studied by \citet{2025A&A...696A.136N}. We investigated the amount of depletion at these two evolutionary stages, the MS and early RGB (when the 1DU is already completed), using their stellar models. In this work, our chemical evolution models assumed that the gas composition has the primordial Li ($\mathrm{A(Li)}=2.69$ dex) and evolves further with contributions from nova explosions in binary systems. 
Together with the corrections from stellar models, we found that the model with a total Li yield of $^{Li}Y_\mathrm{Nova}=2.34\times 10^{-5}\Msun$ produced over the lifetime of a nova, provides the best fit to the observed galactic Li in the thin disc. This constrained yield is in good agreement with the estimated mean value from observations. The model consistently reproduces the Li evolution constructed from both the dwarf-MS and the early RGB stars. 
In this regard, it should be noted that we assumed all novae produce the same amount of Li during their lifetimes. A more complex model, with a dependence of $^{Li}Y_\mathrm{Nova}$ on mass, metallicity or delay timescale might help us to have more precise constraints.

Besides, for the sake of completeness, we note that the only interstellar Li measurement in metal-poor material provides a Li abundance in agreement with that of warm halo stars, suggesting that a different solution from stellar model is required to explain the \say{cosmological Li problem} \citep[][]{2024A&A...690A..38M}. In this regard, we should clarify that our attempt in this work - using the correction relative to the primordial value from stellar models to reproduce the Spite plateau - does not rule out the possibility that the interstellar gas composition has the Li abundance measured in dwarf-MS stars. This possibility requires more thorough investigation and will be addressed in the coming work. 

Finally, we note that \citet{2020MNRAS.496.2902M} studied the Li abundance in the accreted dwarf galaxy Gaia-Sausage-Enceladus, \citet{2010A&A...519L...3M} in $\omega$ Centauri, and \citet{2014MNRAS.444.1812M} in the Sagittarius globular cluster M54. In addition, \citet{2021MNRAS.505..200M} presented chemical evolution models of Li for dwarf spheroidal and ultra-faint galaxies. These authors concluded that the Spite plateau could be a universal feature. This broadens the possibility of testing the internal mixing processes from stellar models, namely, whether they are needed and/or their efficiencies. A comprehensive study on this topic is reserved for future work.

\begin{acknowledgements}
The authors thank the anonymous referee for the kind comments and instructive suggestions. We are also thankful to Paolo Molaro for his insight and thorough suggestions, which have helped improve this paper since the first place. 
This project has received funding from the European Union’s Horizon 2020 research and innovation programme under grant agreement No 101008324 (ChETEC-INFRA). 
CTN and FR acknowledge the support by INAF Mini grant 2024, ``GALoMS – Galactic Archaeology for Low Mass Stars'' (1.05.24.07.02). 
AJK acknowledges support by the Swedish National Space Agency (SNSA). 
GE acknowledges the contribution of the Next Generation EU funds within the National Recovery and Resilience Plan (PNRR), Mission 4 - Education and Research, Component 2 - From Research to Business (M4C2), Investment Line 3.1 - Strengthening and creation of Research Infrastructures, Project IR0000034 – ``STILES - Strengthening the Italian Leadership in ELT and SKA''. 
FR is a fellow of the Alexander von Humboldt Foundation. FR acknowledges support by the Klaus Tschira Foundation. FR and GC acknowledge financial support under the National Recovery and Resilience Plan (NRRP), Mission 4, Component 2, Investment 1.1, Call for tender No. 104 published on 2.2.2022 by the Italian Ministry of University and Research (MUR), funded by the European Union - NextGenerationEU - Project ‘Cosmic POT’ (PI: L. Magrini) Grant Assignment Decree No. 2022X4TM3H by the Italian Ministry of Ministry of University and Research (MUR). 
L.M., and G.C. thank I.N.A.F.
for the 1.05.23.01.09 Large Grant - Beyond metallicity: Exploiting the full POtential of CHemical elements (EPOCH) (ref. Laura Magrini).
A.M., and D.R. acknowledge support from the project "LEGO – Reconstructing the building blocks of the Galaxy by chemical tagging" (P.I. A. Mucciarelli). granted by the Italian MUR through contract PRIN 2022LLP8TK\_001. Supported by Italian Research Center on High Performance Computing Big Data and Quantum Computing
(ICSC), project funded by European Union - NextGenerationEU - and National Recovery and Resilience Plan
(NRRP) - Mission 4 Component 2 within the activities of Spoke 3 (Astrophysics and Cosmos Observations).
 L.M.  thank INAF for the support (Large Grants EPOCH and WST), the Mini-Grants Checs (1.05.23.04.02), and the financial support under the National Recovery and Resilience Plan (NRRP), Mission 4, Component 2, Investment 1.1, Call for tender No. 104 published on 2.2.2022 by the Italian Ministry of University and Research (MUR), funded by the European Union – NextGenerationEU – Project ‘Cosmic POT’ Grant Assignment Decree No. 2022X4TM3H by the Italian Ministry of the University and Research (MUR). 
\end{acknowledgements}

\bibliographystyle{aa}
\bibliography{references}

\end{document}